\documentclass[12pt,preprint,longbibliography,aps, nofootinbib]{revtex4-1}
\usepackage{amsmath, amsthm, amssymb}
\usepackage{graphicx}
\usepackage{subfigure}
\usepackage{wrapfig}
\usepackage{color}

\begin{document}
\title{Absolute Measurement Of Laminar Shear Rate Using Photon Correlation Spectroscopy}
\author{Elliot Jenner}
\author{Brian D'Urso}
\affiliation{Department of Physics \& Astronomy, University of Pittsburgh, Pittsburgh, PA 15260}

\begin{abstract}
An absolute measurement of the components of the shear rate tensor $\mathcal{S}$ in a fluid can be found by measuring the photon correlation function of light scattered from particles in the fluid. Previous methods of measuring $\mathcal{S}$ involve reading the velocity at various points and extrapolating the shear, which can be time consuming and is limited in its ability to examine small spatial scale or short time events. Previous work in Photon Correlation Spectroscopy has involved only approximate solutions, requiring free parameters to be scaled by a known case, or different cases, such as 2-D flows, but here we present a treatment that provides quantitative results directly and without calibration for full 3-D flow. We demonstrate this treatment experimentally with a cone and plate rheometer.
\end{abstract}

\maketitle

\section{Introduction}
Measurement of shear rate in a liquid is of great importance for characterization of fluid and surface slip properties. One method for measuring these properties is to examine the fluid force or torque on an object (e.g. in a rheometer), but this gives an average indication of the slip properties over a large area, and does not directly evaluate the behavior of the fluid. Direct fluid measurements can be taken using hot wire anemometry~\cite{HotWire} and Laser Doppler Velocimetry (LDV)~\cite{LDVReview}, which can give localized and time-resolved measurements, but all of these methods measure fluid velocity. Measuring the shear rate with these methods requires measuring a velocity component at a series of closely spaced points, from which the average slope is deduced; this requires some measurement time at each location and the region of interest must be scanned, so time-resolved measurements of shear are typically not practical with these methods. Furthermore, since hot-wire anemometry requires holding a probe in the flow, it may alter the flow and is sensitive to nearby surfaces~\cite{HotWireNearWall1,HotWireNearWall2}.

In this letter we demonstrate the use of Photon Correlation Spectroscopy (PCS) as a method of measuring the components of $\mathcal{S}$ that does not require any sensor repositioning, measures the shear rate directly and with minimal averaging, does not require calibration or comparison to a reference, and can be performed to within a few percent within minutes. PCS measurements have been used to probe turbulent two-dimensional flow~\cite{dirturbsh}, Brownian motion~\cite{brownian}, and have been used to approximately measure three-dimensional flow in a Couette cell and four-roll mill~\cite{FourBall}. The mathematical treatments in prior work have either been limiting cases (2-D) or have been approximate derivations with free parameters, requiring calibration using a reference. In this letter, we detail our analysis and recent results, which yield rapid PCS measurements of the shear rate with remarkable absolute accuracy in 3D laminar flow and the beginning of secondary flow in a cone and plate rheometer (CPR) with no free parameters.

\section{Theory}
\label{sec:theory}

As with the LDV method, PCS measurements use laser light that is scattered from small particles that seed the flow~\cite{ChuLaser}. The scattered light is collected and the interference between the light scattered from pairs of particles results in the decay of the intensity autocorrelation function. At higher shear rates there is a larger spread of particle velocities and thus a wider spread of Doppler-shifted frequencies in the scattered light, which decreases the decay time of the autocorrelation function. Unlike LDV, which uses a local oscillator (or multiple beams) for a heterodyne measurement, PCS is a homodyne measurement; no additional signal is mixed in.

PCS measures the normalized intensity autocorrelation function $G(\tau)=\frac{\langle I(t)I(t+\tau) \rangle}{|\langle I(t) \rangle|^2} - 1$, where the angular brackets designate an average over time and $I$ indicates the intensity of light scattered from the particles~\cite{Dynamic}. Scattering occurs in a volume containing $N$ scattering particles defined by the intersection of the input beam and the field of view of the detector, which is a collimating lens feeding into a single mode optical fiber. If the input beam is a collimated Gaussian laser beam, then the electric field incident on the $j^{th}$ particle is
\begin{align}
\vec{E}_{\textit{incident},j}(t) &= \vec{E}_0 e^{-i\vec{k}_0 \cdot \vec{r}_j(t)+i\omega_0 t} e^{-[\vec{r}_j(t) \cdot \vec{a}]^2} e^{-[\vec{r}_j(t)\cdot \vec{b}]^2} \nonumber \\
&= \vec{E}_0 e^{-i\vec{k}_0^T \vec{r}_j(t)+i\omega_0 t} e^{-\vec{r}_j(t)^T \mathcal{Z}_1 \vec{r}_j(t)} 
\label{eqn:Einput}
\end{align}
where $\vec{r}_j$ is the displacement from the center of the scattering volume to the $j^{th}$ particle, $\vec{a}=\frac{\hat{a}}{w_{a}}$ and $\vec{b}=\frac{\hat{b}}{w_{b}}$ are the semi-major and semi-minor axes of the input beam (allowing for the possibility of an elliptical beam) where $w_a$ and $w_b$ indicate the radius at which the field drops to $\frac{1}{e}$ in that direction, and $\vec{k}_0$ is the input beam wave vector along the axis of the input beam ($|\vec{k}_0|=\frac{2\pi n}{\lambda}$). We have written this using the tensor $\mathcal{Z}_1 = \vec{a}\vec{a}^T +\vec{b}\hspace{0.5mm}\vec{b}^T$ 
 to simplify and generalize the form of the equations, where all vector quantities are column vectors. For a circular input beam, $w_{a} = w_{b} = w_0$ and $2w_0$ is the conventional intensity $\frac{1}{e^2}$ beam diameter commonly given by laser manufactures.

When light scatters from a particle illuminated by the input beam, the scattered electric field is a spherical wave, and at center of the lens of our detector it can be written as
\begin{equation}
\vec{E}_{\textit{scattered},j}(t)= \vec{E}_{\textit{incident},j}(t) \frac{e^{-i \kappa_{j} |\vec{R}-\vec{r}_j(t)|+i\omega_{sj} t}}{|\vec{R}-\vec{r}_j(t)|}\approx \vec{E}_{\textit{incident},j}(t) \frac{e^{-i \vec{\kappa}_{j}\cdot\vec{R}}}{R} e^{i \vec{\kappa}_{j}\cdot\vec{r}_j(t)}e^{i\omega_{sj} t}
\label{eq:Es}
\end{equation}
where $\vec{R}$ is the displacement from the center of the scattering volume to the center of the detector along the optical axis of the detector, $\omega_{sj}$ is the frequency of the light scattered off the $j^\textrm{th}$ particle and $\vec{\kappa}_{j}$ is the scattering vector from the  $j^\textrm{th}$ particle ($|\vec{\kappa}_j|=\frac{2\pi n}{\lambda}$; $n$ is the fluid index of refraction and $\lambda$ is the laser wavelength). Looking at the equations, we see that $\vec{\kappa}_{j} \parallel (\vec{R}-\vec{r}_j(t))$, so $\kappa_{j} |\vec{R}-\vec{r}_j(t)| = \vec{\kappa}_{j} \cdot (\vec{R}-\vec{r}_j(t))= \vec{\kappa}_{j}\cdot\vec{R}-\vec{\kappa}_{j}\cdot\vec{r}_j(t)$. We have also taken the approximation that $\vec{R} \gg \vec{r}_j(t)$ in the final form of the denominator of Eq.~\ref{eq:Es}.

Allowing for the possibility of an elliptical detector, the lens-fiber detector system has effective major and minor axes $\vec{c}=\frac{\hat{c}}{w_{c}}$ and $\vec{d}=\frac{\hat{d}}{w_{d}}$ respectively; $w_c$ and $w_d$ indicate the radius at which the electric field of a beam emitted from the fiber through the lens drops to $\frac{1}{e}$ in that direction. Only a fraction of the light scattered off each particle is coupled in the single mode fiber by the lens of the detector. Thus the field in the center of  the fiber is $\vec{E}_j(t) \propto \sqrt{\eta_j(t)}\vec{E}_{\textit{scattered},j}(t)$, where $\eta_j(t)$ is the coupling efficiency, the fraction of the power transferred to the detector by the light scattered from the $j^{th}$ particle. 
We find that for coupling of a spherical wave to a fiber using a collimating lens, treated with a TEM$_{00}$ Gaussian mode, $\eta_j(t) \propto |e^{-(\vec{r}_j(t) \cdot \vec{c})^2}e^{-(\vec{r}_j(t)\cdot \vec{d})^2}|^2 = |e^{-\vec{r}^T_j(t) \mathcal{Z}_2 \vec{r}_j(t)}|^2$, where $\mathcal{Z}_2 = \vec{c}\hspace{0.5mm}\vec{c}^T +\vec{d}\vec{d}^T$ (See Appendix \ref{sec:Coupling}).\cite{Coupling} The spatial distribution of the coupling was not considered in previous work.\cite{DiffusionShear,dirturbsh} Then, the electric field in center of the fiber from a single scattering particle is
\begin{equation}
\vec{E}_j(t) \propto \frac{e^{-i \vec{\kappa}_{j}^T \vec{R}}}{R} \vec{E}_{\textit{incident},j}(t) e^{i \vec{\kappa}_{j}^T \vec{r}_j(t)+i\omega_{sj} t} e^{-\vec{r}^T_j(t) \mathcal{Z}_2 \vec{r}_j(t)}
 \label{eq:Et}
\end{equation}

Combining Eq.~\ref{eq:Et} and Eq.~\ref{eqn:Einput}, we find that the detected field from a single scattering particle is
\begin{equation}
\vec{E}_j(t)=\vec{E}_{D} e^{-i \vec{k}_j^T \vec{r}_j(t)+i\omega'_j t} e^{-\vec{r}_j^T(t) \mathcal{Z} \vec{r}_j(t)}
\label{eq:E}
\end{equation}
where $\omega'_j=\omega_0+\omega_{sj}$ is the Doppler-shifted frequency of the light scattered off the $j^\textrm{th}$ particle, $\vec{k}_j= \vec{k}_0-\vec{\kappa}_j$ is the net scattering vector ($|\vec{k}_j|=\frac{4\pi n}{\lambda} \sin(\theta_j/2)$; $\theta_j$ is the scattering angle for the $j^\textrm{th}$ particle), $\mathcal{Z} = \mathcal{Z}_1 + \mathcal{Z}_2$, and $\vec{E}_{D} \propto \vec{E}_{0}\frac{e^{-i \vec{\kappa}_{j}^T \vec{R}}}{R}$ is the magnitude of the electric field which is constant. Because of the form of the equation for $G(\tau)$, its value is independent of $\vec{E}_{D}$ .

We assume that the particles are small enough that the velocity of each particle is the same as the velocity of the fluid at that location, that the fluid velocity is time-independent during the measurement, and that beam is sufficiently small that the velocity of the fluid can be described by a first order expansion about the center of the beam. So the position of the particles at some later time $t+\tau$ can be written as
\begin{equation}
\vec{r}_j(t+\tau)=\vec{r}_{j}(t)+\vec{v}\tau+\mathcal{S}\vec{r}_{j}(t)\tau + \vec{r}_{\textit{dif},j}(\tau)
\label{eq:rt}
\end{equation}
where $\vec{v}$ is the average velocity of the fluid, $\vec{r}_{\textit{dif},j}(\tau)$ is additional displacement due to diffusion in the time $\tau$ between $t$ and $t+\tau$, and $\mathcal{S}$ is the shear rate tensor with components $\mathcal{S}_{{\alpha}{\beta}} = \frac {\partial v_{\alpha}} {\partial r_{\beta}}$. Then at a time $t+\tau$ Eq.~\ref{eq:E} becomes
\begin{equation}
\vec{E}_j(t+\tau)=\vec{E}_D e^{-i \vec{k}_j^T [\vec{r}_{j}(t)+\vec{v}\tau+\mathcal{S}\vec{r}_{j}(t)\tau + \vec{r}_{\textit{dif},j}(\tau)]+ i\omega'_j (t+\tau)} e^{-[\vec{r}_{j}(t)+\vec{v}\tau]^T \mathcal{Z} [\vec{r}_{j}(t)+\vec{v}\tau]} 
\end{equation}

Returning to the definition of $G(\tau)$, we have 
\begin{equation}
G(\tau) =\frac{\langle I(t)I(t+\tau) \rangle}{\langle I(t) \rangle^2} - 1 = \frac{\sum\limits_{ijkl}\left\langle  E_i(t)E^*_j(t)E_k(t+\tau)E_l^*(t+\tau) \right\rangle}{\left|\sum\limits_{mn}\left\langle E_m(t)E_n^*(t) \right\rangle \right|^2} - 1
\end{equation}
Terms relating more than two particles do not survive averaging over time if the particle distribution is random and the particles are spread over a volume larger than the wavelength of light used in all three dimensions~\cite{pairwiseonly}. Applying this condition, only terms with pairs of particles can survive in the numerator. We also note that the expectation values of the different sums are independent. Therefore,
\begin{align}
G(\tau) &= \frac{\sum\limits_{ik}\left\langle  E_i(t)E^*_i(t)E_k(t+\tau)E_k^*(t+\tau) \right\rangle+\sum\limits_{ik}\left\langle  E_i(t)E^*_k(t)E_k(t+\tau)E_i^*(t+\tau) \right\rangle}{\left|\sum\limits_{mn}\left\langle E_m(t)E_n^*(t) \right\rangle \right|^2} - 1 \nonumber \\
&= \frac{\left|\sum\limits_{j}\left\langle E_j(t)E^*_j(t)\right\rangle\right|^2+\left|\sum\limits_{j}\left\langle  E_j(t)E_j^*(t+\tau) \right\rangle\right|^2}{\left|\sum\limits_{mn}\left\langle E_m(t)E_n^*(t) \right\rangle \right|^2} - 1
\end{align}

For particles inside the scattering volume, there will be very little variation in the scattering vectors, so we can let $\vec{k}_j \approx \vec{k} = \vec{k}_0 - \vec{\kappa}$, where $\vec{k}$ is the average total scattering vector and $\vec{\kappa}$ is the average scattered wave vector (along the axis of the detector). We see that in the sums each term is identical in form and integrated over all space, so we have the sum of $N$ identical terms. We also note that displacement due to diffusion is independent of the particle position, and so is integrated separately, and is probabilistic with a distribution $P(\vec{r}_{\textit{dif},j}(\tau))=\frac{1}{\sqrt{4\pi D \tau}^n} e^{-\frac{\vec{r}^2}{4D\tau}}$, where $D$ is the diffusion constant and $n$ is the dimensionality of the motion \cite{einsteinBrownian,brownian}. Terms in the denominator with $m\neq n$ are zero due to the properties of Gaussian integrals, and thus the first term in the numerator and the denominator are constant and equal. The second term in the numerator can be easily integrated (see Appendix \ref{sec:Expectation}). Therefore 
\begin{equation}
G(\tau)= G_0 e^{-2Dk^2\tau}e^{-\vec{v}^T \mathcal{Z}\vec{v} \tau^2}e^{- \frac{1}{4}\vec{k}^T \mathcal{S} \mathcal{Z}^{-1} \mathcal{S}^T \vec{k}\tau^2}
\label{eqn:Gfinal}
\end{equation}
where $D = \frac{k_B T}{3\pi \eta d}$ is the diffusion constant, $k_B$ is Boltzmann's constant, $T$ is the temperature of the liquid, $d$ is the diameter of the scattering particles, and $\eta$ here is the dynamic viscosity of the fluid. The constant $G_0$ is included to account for detection efficiency considerations, minimum measurement time and noise; its exact value is unimportant. Note that the measurement is not sensitive to components of $\mathcal{S}$ involving the velocity perpendicular to the plane containing $\vec{k}$, $\vec{k}_0$ and $\vec{\kappa}$.

The average velocity is important only at very high velocities ($\vec{v}^T \mathcal{Z}\vec{v} \gg \langle \mathcal{S}_{\alpha\beta} \rangle^2$)~\cite{dirturbsh}, where particles will be pass into and out of the beam during the course of a measurement. We can write the diffusion factor as $e^{-q\tau}$, where $q=2Dk^2$ in the case of no shear; in the presence of shear, $q$ also includes effects such as Taylor Diffusion~\cite{DiffusionShear}. Writing the factor in this way allows us to easily account for the effect of diffusion while fitting our data without worrying about the precise value of $q$ under given conditions. This effective diffusion enhancement is visible in all data, explaining the initial non-zero slope of $G(\tau)$, but does not effect the shear measurement. The value of $G_0$ is obtained during fitting but is not relevant for determining the shear rate. Therefore we may write Eq.~\ref{eqn:Gfinal} as
\begin{equation}
G(\tau)= G_0 e^{-q\tau} e^{- \frac{1}{4}\vec{k}^T \mathcal{S} \mathcal{Z}^{-1} \mathcal{S}^T \vec{k}\tau^2}
\label{eqn:GFinalCorrected}
\end{equation}
which can be easily fit once the tensors in the exponent are evaluated for the geometry at hand. In contrast to previous theoretical treatment~\cite{FourBall}, Eq.~\ref{eqn:GFinalCorrected} can be quantitatively fit without the need for scaling or calibration, as the unknown factors ($q$ and $G_0$) are unimportant and all other factors in the equation can be determined independently. We note that this formalism returns the previously reported 2-D equations\cite{dirturbsh}, with corrections for the inclusion of the coupling effect, when $\mathcal{S}$, $\mathcal{Z}$ and $\vec{k}$ are written for that case (ie appropriate components are set to $0$).

\section{Experimental}

Laminar flow is well understood in a CPR, so it is easy to predict the results of the shear rate measurements. We use cylindrical coordinates ($\rho, \phi, z$) with the origin at the intersection of the axis of rotation of the cone and the surface of the flat plate and $\hat{z}$ parallel to the rotation axis of the cone. In laminar flow, if the angle between the cone surface and the horizontal is $\alpha \lesssim 6^\circ$, it can be shown, using the continuity equation and no slip boundary conditions, that there is only a single non-zero tensor component $s=\frac{\partial v_{\phi}}{\partial z}=\omega/\alpha$; $\omega$ is the angular frequency of the cone about the vertical ($z$) axis\cite{rheology}. In this limit Eq.~\ref{eqn:GFinalCorrected} reduces to

\begin{equation}
G(\tau)= G_0 e^{-q\tau}e^{-\frac{k_{\phi}^2 s^2 w_b^2 w_d^2}{4(w_b^2 + w_d^2)} \tau^2}
\label{eq:GLaminar}
\end{equation}

Where $k_\phi = \vec{k}\cdot \hat{\phi}$. Note that $G(\tau)$ in this case depends only on the size of the beam and the receiver in the $\hat{z}$ direction. For laminar flow we can simply fit Eq.~\ref{eq:GLaminar} to find $\frac{k_{\phi}^2 s^2 w_b^2 w_d^2}{4(w_b^2 + w_d^2)}$, which has no unknown parameters except $s$
. In the event we leave laminar flow, other additive terms in the exponent may appear; we do not investigate this behavior in this letter.

To measure the shear tensor, we direct a Gaussian laser beam from a HeNe laser ($6$~mW, $\lambda=632.8$~nm, beam diameter $2w=0.81$~mm) into a cylindrical glass tank and through the cone-plate gap of our CPR. The tank is filled with water in which we have suspended a low concentration of spheres of diameter d=0.4~$\mu$m, ultrasonically dispersed, as scattering particles (they are small enough to follow the local velocity; the Stokes number is much less than unity~\cite{snyder1971}). The concentration of particles is kept low to avoid multiple scattering. Typically, cone and plate rheometers are used with liquid only in the gap between the cone and plate~\cite{rheology}; to minimize errors caused by the excess water outside of the gap, we avoid taking measurements close to the edge of the cone. With this incident intensity the photon counting rate is of the order of $750$~kHz to $1.5$~MHz (when shielded from ambient light) at a scattering angle of $20.5^\circ$. The measurement count rate will increase if unshielded, but as this light will be uncorrelated the measurement of shear is insensitive to ambient light levels. We drive the cone with a stepper motor, and take measurements by orienting the beam and receiver in the horizontal plane, parallel to the plate. The collimating lens receives the scattered light, which is coupled into the single-mode optical fiber to an avalanche photodiode based single photon counting module (SPCM), so that the direction of the received light is well defined. The SPCM signal is fed into an ALV-5000 autocorrelation card that converts the intensity $I(t)$ into a (normalized) autocorrelation function $G(\tau)$. The effective sizes $w_c$ and $w_d$ of the detector can be found by measuring the dimensions of a laser beam sent backward through the fiber and out through the collimator. 

\begin{figure}[!tbh]
\centering
\includegraphics[width=0.50\textwidth]{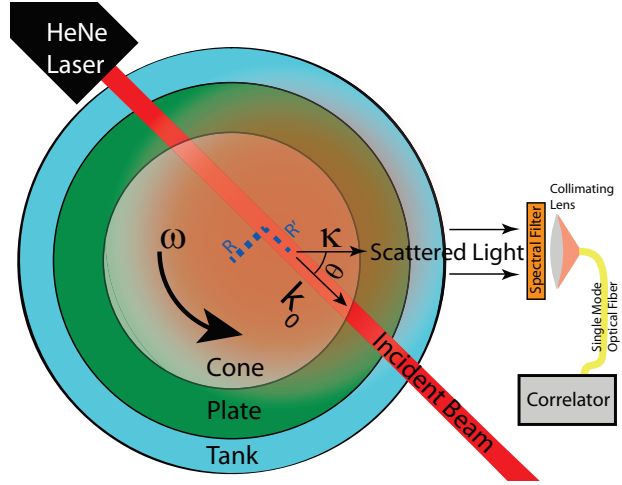}
\caption{System geometry top view. The incident beam scatters light over a wide angular range but only one scattering vector $\vec{\kappa}$ is collected by the single mode fiber via the collimating lens. The light output is fed into an avalanche photodiode photon counter, then the correlation card. The correlation yields the (time-averaged) shear rate $s$.}
\label{fig:pcs_setup} 
\end{figure}


\section{Results and Discussion}

\begin{figure*}[!tbh]
\centering
\includegraphics[width=0.99\textwidth]{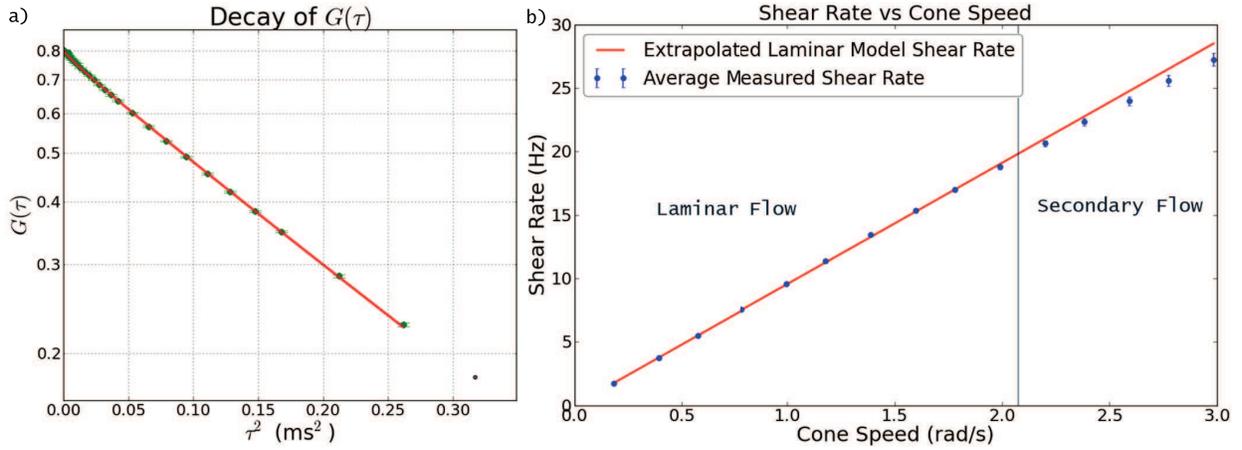}
    \caption{a) Data and fit for a measurement taken at 0.4~rad/s cone speed at a scattering angle of 20.5$^\circ$, with a $2w_0=0.81~mm$ beam and a $2w_s = 1.05~mm$ detector (on a logarithmic scale vs $\tau^2$, so $G(\tau)$ is a straight line except near $\tau^2 = 0$). b) The measured and theoretical shear rate as a function of cone speed at a scattering angle of 20.5$^\circ$ for a single run.}            
\label{fig:data}
\end{figure*}

An example of PCS data and a fit curve to it is shown in Fig.~\ref{fig:data}a. An example of fitting shear rate vs cone speed and a comparison of the measured shear rate to the theoretical shear rate in the laminar limit is shown in Fig.~\ref{fig:data}b. Secondary flow begins at $\omega\approx 2$~rad/s and turbulence begins at $\omega\approx 15$~rad/s~\cite{C&Psecondary}. Data sets were gathered by increasing the cone speed from 0.2~rad/s to 3.0~rad/s in increments of 0.2~rad/s, with data at each cone speed averaged from 20 repeated readings at that speed with an averaging time of 10~s each. Five complete data sets were gathered, with a full removal of the sample and realignment between sets, to test the repeatability of the measurement. 

We see remarkably good agreement with the theoretical values of shear rate in the laminar regime, where the average slope of the Shear Rate vs Cone Speed graph over the 5 data sets was $9.61 \pm 0.03$~$rad^{-1}$, a $0.6\%$ error compared to the theoretical slope of $9.55$~$rad^{-1}$. Slopes from individual data sets showed a statistical uncertainty of at most $\pm 0.5\%$, while individual slopes varied from the theoretical value by a maximum of $\pm 2\%$, larger than could be accounted for by statistical error alone, indicating additional sources of error in our experiment.

Tests of repeatability indicate that the additional error results from limited repeatability in assembly and alignment, including the setting of the scattering angle, measurement of the widths of the input beam and the lens/fiber detector, determination of the direction of the velocity at the measurement point (which is needed to properly perform the scalar product $\vec{k} \cdot \hat{\phi}$), errors vertically aligning the beam and the receiver with each other and the gap, and error in the distance of the cone from the plate. Despite these issues, our measurements were still highly accurate, indicating that the method is quite robust.

\section{Conclusion}
We have demonstrated the efficacy of Photon Correlation Spectroscopy for measuring the shear rate in a three dimensional system, with a short individual measurement time. We show that this measurement is quantitatively precise, removing the need to scale to a reference found in some previous work, and tolerant of minor systematic errors. By measuring the shear rate directly and rapidly, we open up the possibility of looking at phenomena that cannot be examined utilizing slower methods or methods that depend on averaging over large areas.

\section{Acknowledgments}
We would like to thank Dr. Walter Goldburg (University of Pittsburgh) for his invaluable assistance in this work, as well as for allowing us to utilize some of his equipment for measurements.

\appendix

\section{Coupling Efficiency}
\label{sec:Coupling}
In order to determine the coupling efficiency $\eta_j(t)$ , we rewrite Eq.~\ref{eq:Es} for any observation point. 
\begin{equation}
\vec{E}_{\textit{scattered},j}(\vec{h},t)= \vec{E}_{\textit{incident},j}(t) \frac{e^{-i K |\vec{h}+\vec{g}(t)|}}{|\vec{h}+\vec{g}(t)|}
\end{equation}
where $K$ is the magnitude of the scattering vector parallel to $\vec{h}+\vec{g}(t)$ ($|\vec{K}|=\frac{2\pi n}{\lambda}$), $\vec{g}(t) = \vec{R}-\vec{r}_j(t)$ is the displacement from the particle to the center of the collimating lens, and $\vec{h}$ is the vector to the observation point from the center of the lens. We neglect the phase factor $e^{i\omega_s t}$ as it has no spatial dependence. Although the vector $\vec{K}$ changes direction as a function of $\vec{h}$, the magnitude is constant.

The field in a plane parallel to the plane of the lens may be represented by the collimated Gaussian TEM$_{00}$ mode
\begin{equation}
E_f(\vec{h}) = E_{fi} e^{-(\vec{h} \cdot \vec{c})^2} e^{-(\vec{h} \cdot \vec{d})^2}
\end{equation}
where $E_{fi}$ is the field magnitude.

The coupling efficiency is given by the overlap integral
\begin{equation}
\eta_j(t)=\frac{\left|\int\limits_A E_{\textit{scattered},j}(\vec{h},t) E_f^*(\vec{h}) dA\right|^2}{\left|\int\limits_A E^2_{\textit{scattered},j}(\vec{h},t) dA\right| \left|\int\limits_A E_f^2(\vec{h}) dA\right|}
\end{equation}
which indicates the ratio of power coupling between the modes through a surface $A$ to the total power in each mode~\cite{Coupling}. Because $G(\tau)$ depends on the ratio of intensities, in all cases the constant normalization terms in the denominator are unimportant.

Turning to the coupling term, for the plane containing the lens we have 
\begin{equation}
\eta_j(t) \propto \left|\int\limits_A E_{\textit{scattered},j}(\vec{h},t) E_f^*(\vec{h}) dA\right|^2 \approx \left|\int\limits_A \frac{e^{-i K |\vec{h}+\vec{g}(t)|}}{|\vec{h}+\vec{g}(t)|}  e^{-(\vec{h} \cdot \vec{c})^2} e^{-(\vec{h} \cdot \vec{d})^2} dA\right|^2
\label{EtaAtLens}
\end{equation}

which is non-analytic. However, since the integral is over this infinite plane, the integral must be the same on any parallel surface (so long as the reversed detector beam would still be approximately collimated when it passes through that surface), and we are free to choose the most convenient one. Thus we evaluate that integral on the surface containing the scattering particle, where $E_{\textit{scattered},j}(\vec{h},t)=E_{\textit{incident},j}(t) \delta(\vec{h}-\vec{g}(t))$, which gives

\begin{equation}
\eta_j(t) \propto \left| e^{-(-\vec{g}(t) \cdot \vec{c})^2} e^{-(-\vec{g}(t) \cdot \vec{d})^2}\right|^2
\label{eq:EtaAtPart} 
\end{equation}

We can also approximately evaluate the integral on the plane containing the lens. Looking at the spherical wave, $|\vec{h}+\vec{g}(t)| = \sqrt{h^2+g(t)^2+2\vec{h}\cdot\vec{g}(t)}\approx g(t)+\frac{h^2}{2g(t)}+\frac{\vec{h}\cdot\vec{g}(t)}{g(t)}$, where we have Taylor expanded to $1^{st}$ order with $g(t)>h$; we cannot a priori say that $h^2/2 \ll \vec{h} \cdot \vec{g}(t)$ (as $\vec{g}(t)$ and $\vec{h}$ are nearly perpendicular). If we retain both terms in this expansion, $E_{\textit{scattered},j}(\vec{h},t) \approx E_{\textit{incident},j}(t) \frac{e^{-i K \left(g(t)+\frac{h^2}{2g(t)}+\frac{\vec{h}\cdot\vec{g}(t)}{g(t)}\right)}}{g(t)}$, so 
\begin{align}
\eta_j(t) &\propto  \left| \frac{e^{-i K g(t)}}{g(t)}\int  e^{-i K\left(\frac{h^2}{2g(t)}+\frac{\vec{h}\cdot\vec{g}(t)}{g(t)}\right)}  e^{-(\vec{h} \cdot \vec{c})^2} e^{-(\vec{h} \cdot \vec{d})^2} dA\right|^2 \nonumber \\
&= \left| \frac{e^{-i K g(t)}}{g(t)} \frac{1}{\sqrt{(\frac{2}{w_c^2}-i\frac{K}{g(t)})(\frac{2}{w_d^2}-i\frac{K}{g(t)})}} e^{-\frac{K^{2} (-\vec{g}(t)\cdot \vec{c})^2 w_c^4}{4g(t)^2-2ig(t)K w_c^2}}e^{-\frac{K^{2} (-\vec{g}(t)\cdot \vec{d})^2 w_d^4}{4g(t)^2-2ig(t)K w_d^2}}\right|^2
\label{eq:EtaApprox}
\end{align}

If we do not retain $h^2/2$, then we have performed the plane wave approximation and $E_{\textit{scattered},j}(\vec{h},t) \approx E_{\textit{incident},j}(t) \frac{e^{-i K \left(g(t)+\frac{\vec{h}\cdot\vec{g}(t)}{g(t)}\right)}}{g(t)}$, which gives
\begin{align}
\eta_j(t) &\propto \left| \frac{e^{-i K g(t)}}{g(t)}\int  e^{-iK\frac{\vec{h}\cdot\vec{g}(t)}{g(t)}}  e^{-(\vec{h} \cdot \vec{c})^2} e^{-(\vec{h} \cdot \vec{d})^2} dA\right|^2 \nonumber \\
&= \left| \frac{e^{-i K g(t)}}{g(t)} \pi w_c w_d e^{-\frac{K^{2}}{4g(t)^2}[(-\vec{g}(t)\cdot \vec{c})^2 w_c^4+(-\vec{g}(t)\cdot \vec{d})^2 w_d^4]}\right|^2
\label{eq:EtaVeryApprox}
\end{align}

\begin{figure*}[!tbh]
\centering
\includegraphics[width=0.99\textwidth]{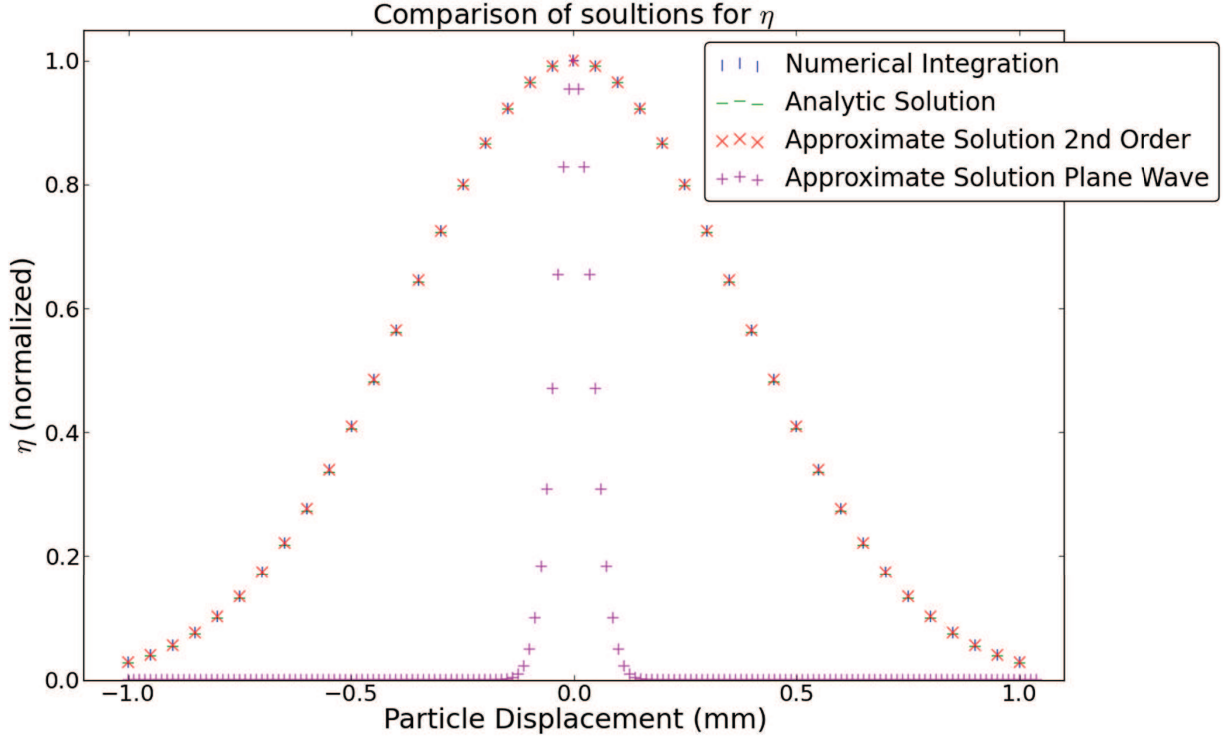}
    \caption[Comparison of the coupling efficiency as calculated by different methods]{Comparison of the coupling efficiency in water at $g=200$~mm and $\lambda = 633$~nm, as calculated numerically on the plane of the lens, analytically evaluated on the plane containing the particle, and approximately integrated on the plane of the lens. We also show the result of evaluating with a plane wave approximation (neglecting $h^2/2$) to demonstrate that this is an overapproximation.}
\label{fig:overlap}
\end{figure*}

A comparison of Eq.~\ref{eq:EtaAtPart}, Eq.~\ref{eq:EtaApprox} and the numerical integral of Eq.~\ref{EtaAtLens} shows excellent agreement when $g$ is small compared to $K w_c$ and $K w_d$, which is precisely the limit where diffraction effects are not large. By contrast, Eq~\ref{eq:EtaVeryApprox} is very different, showing that the plane wave approximation is a poor choice in this case (See Fig.~\ref{fig:overlap}).

Finally, using the exact solution (Eq.~\ref{eq:EtaAtPart}), remembering that $\vec{g}(t) = \vec{R}-\vec{r}_j(t)$, and noticing that $\vec{R} \perp \vec{c}$ and $\vec{R} \perp \vec{d}$ by definition, we see that
\begin{equation}
\eta_j(t) \propto \left|e^{-(-[\vec{R}-\vec{r}_j(t)] \cdot \vec{c})^2}e^{-(-[\vec{R}-\vec{r}_j(t)] \cdot \vec{d})^2}\right|^2=\left|e^{-(\vec{r}_j(t) \cdot \vec{c})^2}e^{-(\vec{r}_j(t) \cdot \vec{d})^2}\right|^2 = |e^{-\vec{r}^T_j(t) \mathcal{Z}_2 \vec{r}_j(t)}|^2
\end{equation}
where $\mathcal{Z}_2 = \vec{c}\hspace{0.5mm}\vec{c}^T +\vec{d}\vec{d}^T$.

\section{Evaluation Of G($\tau$)}
\label{sec:Expectation}
The normalized correlation function is
\begin{equation}
G(\tau) =\frac{\langle I(t)I(t+\tau) \rangle}{|\langle I(t) \rangle|^2} - 1 = \frac{\sum\limits_{ijkl}\left\langle E_i(t)E^*_j(t)E_k(t+\tau)E_l^*(t+\tau) \right\rangle}{\left|\sum\limits_{mn}\left\langle E_m(t)E_n^*(t) \right\rangle\right|^2} - 1
\end{equation}

Terms relating more than two particles do not survive averaging over time if the particle distribution is random and the particles are spread over a volume larger than the wavelength of light used in all dimensions~\cite{pairwiseonly}. Applying this condition, only terms with pairs of particles can survive in the numerator. We also note that the expectation values of the different sums are independent. Therefore,
\begin{align}
G(\tau) &= \frac{\sum\limits_{ik}\left\langle  E_i(t)E^*_i(t)E_k(t+\tau)E_k^*(t+\tau) \right\rangle+\sum\limits_{ik}\left\langle  E_i(t)E^*_k(t)E_k(t+\tau)E_i^*(t+\tau) \right\rangle}{\left|\sum\limits_{mn}\left\langle E_m(t)E_n^*(t) \right\rangle \right|^2} - 1 \nonumber \\
&= \frac{\left|\sum\limits_{j}\left\langle E_j(t)E^*_j(t)\right\rangle\right|^2+\left|\sum\limits_{j}\left\langle  E_j(t)E_j^*(t+\tau) \right\rangle\right|^2}{\left|\sum\limits_{mn}\left\langle E_m(t)E_n^*(t) \right\rangle \right|^2} - 1
\label{eq:G}
\end{align}

For the denominator,
\begin{align}
\left|\langle I(t) \rangle\right|^2 &= \left|\sum\limits_{mn}\langle E_{m}(t)E_{n}^*(t)\rangle\right|^2 = \left|\sum\limits_{m\neq n}\langle E_{m}(t)E_{n}^*(t)\rangle\right|^2 + \left|\sum\limits_{j}\langle E_{j}(t)E_{j}^*(t)\rangle\right|^2\nonumber \\
&= \left|\sum\limits_{j}\langle E_{j}(t)E_{j}^*(t)\rangle\right|^2 =\left| E_D^2 \sum\limits_{j} \langle e^{-2\vec{r}_j^T(t) \mathcal{Z} \vec{r}_j(t)}\rangle\right|^2\nonumber \\
&= \left|E_D^2 N \int\limits_{-\infty}^{\infty}  e^{-2\vec{r}^T(t) \mathcal{Z} \vec{r}(t)} d^nr(t)\right|^2 = \left|E_D^2 N \sqrt{\frac{\pi^n}{det(2\mathcal{Z})}}\right|^2 = E_D^4 N^2 \frac{\pi^n}{det(2\mathcal{Z})}
\label{eq:denom}
\end{align}
where each of the $N$ expectation values is identical and the integral is $n$-Dimensional for generality.\footnote{$\protect\int_{-\infty}^\infty e^{-x^{T} \mathcal{A} x } \, d^nx=\sqrt{\frac{(\pi)^n}{\det \mathcal{A}}}$. This can be verified by applying the unitary transform $a^\dagger a = 1$, where this diagonalizes $\mathcal{A}$, and then performing the integral in eigenspace. Such a transformation does not change the determinant.}

The first term in the numerator of Eq.~\ref{eq:G} is identical to Eq.~\ref{eq:denom}. For particles inside the scattering volume, there will be very little variation in the scattering vectors, so we can let $\vec{k}_j = \vec{k}_0 - \vec{\kappa}_j \approx \vec{k}_0 - \vec{\kappa} = \vec{k}$, where $\vec{k}$ is the average total scattering vector ($\vec{\kappa}$ is the average scattered wave vector which is along the axis of the detector) and $\omega'_j \approx \omega'$, where $\omega'$ is the average total phase shift. We see that each term in the sums is identical in form and integrated over all space. We also note that displacement due to diffusion is independent of the particle position, and so is integrated separately. Therefore 
\begin{align}
&\left|\sum\limits_{j} \langle E_{j}(t)E_{j}^*(t+\tau)\rangle\right|^2 =\nonumber\\
&|E_D^2 \sum\limits_{j} \langle e^{i \vec{k}^T \vec{v}\tau} e^{-\vec{v}^T \mathcal{Z} \vec{v}\tau^2} e^{-i\omega' \tau}  e^{i \vec{k}^T \vec{r}_{\textit{dif},j}(\tau)}e^{i \vec{k}^T\mathcal{S}\vec{r}_j(t)\tau} e^{-2\vec{r}_j^T(t) \mathcal{Z} \vec{r}_j(t)} e^{-2\vec{v}^T \mathcal{Z} \vec{r}_j(t)\tau} \rangle|^2= \nonumber \\
&E_D^4  e^{-2\vec{v}^T \mathcal{Z} \vec{v}\tau^2} \times \nonumber\\
&\;\;\;\;\; \left|\sum\limits_{j} \!\!\! \int\limits_{-\infty}^{\infty} \!\!\!P(\vec{r}_{\textit{dif},j}(\tau))e^{i \vec{k}^T \vec{r}_{\textit{dif},j}(\tau)}d^nr_{\textit{dif},j}(\tau) \!\!\!\int\limits_{-\infty}^{\infty} \!\!\! e^{i \vec{k}^T\mathcal{S}\vec{r}_j(t)\tau} e^{-2\vec{r}_j^T(t) \mathcal{Z} \vec{r}_j(t)} e^{-2\vec{v}^T \mathcal{Z} \vec{r}_j(t)\tau} d^n r_j(t)\right|^2 
\label{eq:2nd}
\end{align}
where $ P(\vec{r}_{\textit{dif}}(\tau))$ is the $n$-D probability distribution of the displacement due to diffusion. The integrals are evaluated over all possible particle positions and diffusion displacements.

For the integral over particle positions in Eq.~\ref{eq:2nd}, we have\footnote{$\protect\int_{-\infty}^\infty e^{-\vec{x}^T \mathcal{A} \vec{x}+\vec{B}^T \vec{x}} d^nx= \sqrt{ \frac{(\pi)^n}{\det{\mathcal{A}}} }e^{\frac{1}{4}\vec{B}^{T}\mathcal{A}^{-1}\vec{B}}$. This can be verified by applying the unitary transform $a^\dagger a = 1$, where this diagonalizes $\mathcal{A}$, and then performing the integral in eigenspace. Such a transformation does not change the determinant.}
\begin{align}
&\int\limits_{-\infty}^{\infty}  e^{i \vec{k}^T\mathcal{S}\vec{r}_j(t)\tau} e^{-2\vec{r}_j^T(t) \mathcal{Z} \vec{r}_j(t)} e^{-2\vec{v}^T \mathcal{Z} \vec{r}_j(t)\tau} d^n r_j(t)= \nonumber \\
&\;\;\;\;\;\; \sqrt{\frac{\pi^n}{det(2\mathcal{Z})}} e^{-\frac{1}{8}\vec{k}^T\mathcal{S} (2\mathcal{Z})^{-1} \mathcal{S}^T \vec{k}\tau^2} e^{-\frac{1}{2}i\vec{k}^T\mathcal{S}\vec{v}\tau^2}e^{\frac{1}{2}\vec{v}^T\mathcal{Z}\vec{v}\tau^2}
\end{align}

The $n$-D probability density for diffusion is $P(\vec{r}(\tau)) = \frac{1}{\sqrt{4\pi D \tau}^n} e^{-\frac{\vec{r}^2}{4D\tau}}$, where $D = \frac{k_B T}{3\pi \eta d}$ is the diffusion constant; $k_B$ is Boltzmann's constant, $T$ is the temperature of the liquid, $d$ is the diameter of the scattering particles, and $\eta$ here is the dynamic viscosity of the fluid~\cite{einsteinBrownian,brownian}. So for the integral over the diffusion displacements in Eq.~\ref{eq:2nd} we have
\begin{align}
&\int\limits_{-\infty}^{\infty} P(\vec{r}_{\textit{dif}_j}(\tau))e^{i \vec{k}^T \vec{r}_{\textit{dif},j}(\tau)} d^nr_{\textit{dif},j}(\tau) = \nonumber \\ 
&\;\;\;\;\;\;\frac{1}{\sqrt{4\pi D \tau}^n} \int\limits_{-\infty}^{\infty} e^{i \vec{k}\cdot \vec{r}_{\textit{dif},j}(\tau)} e^{-\frac{\vec{r}_{\textit{dif},j}(\tau)^2}{4D\tau}} d^nr_{\textit{dif},j}(\tau)
= e^{-Dk^2\tau}
\end{align}

Putting these into Eq.~\ref{eq:2nd}, we have (noting that the $N$ terms in the summation are identical)
\begin{equation}
\left|\sum\limits_{j} \langle E_{j}(t)E_{j}^*(t+\tau)\rangle\right|^2 =E_D^4  N^2 \frac{\pi^n}{det(2\mathcal{Z})} e^{-2Dk^2\tau} e^{-\vec{v}^T \mathcal{Z} \vec{v}\tau^2} e^{-\frac{1}{4}\vec{k}^T\mathcal{S} \mathcal{Z}^{-1} \mathcal{S}^T \vec{k}\tau^2} 
\end{equation}

Therefore,
\begin{equation}
G(\tau) = \frac{\langle I(t)I(t+\tau)\rangle}{|\langle I(t)\rangle|^2} - 1 = e^{-2Dk^2\tau} e^{-\vec{v}^T \mathcal{Z} \vec{v}\tau^2} e^{-\frac{1}{4}\vec{k}^T\mathcal{S}\mathcal{Z}^{-1} \mathcal{S}^T \vec{k}\tau^2}
\label{eqn:appGfinal}
\end{equation}


\begin{thebibliography}{19}
\expandafter\ifx\csname natexlab\endcsname\relax\def\natexlab#1{#1}\fi
\expandafter\ifx\csname bibnamefont\endcsname\relax
  \def\bibnamefont#1{#1}\fi
\expandafter\ifx\csname bibfnamefont\endcsname\relax
  \def\bibfnamefont#1{#1}\fi
\expandafter\ifx\csname citenamefont\endcsname\relax
  \def\citenamefont#1{#1}\fi
\expandafter\ifx\csname url\endcsname\relax
  \def\url#1{\texttt{#1}}\fi
\expandafter\ifx\csname urlprefix\endcsname\relax\def\urlprefix{URL }\fi
\providecommand{\bibinfo}[2]{#2}
\providecommand{\eprint}[2][]{\url{#2}}

\bibitem[{\citenamefont{Comte-Bellot}(1976)}]{HotWire}
\bibinfo{author}{\bibfnamefont{G.}~\bibnamefont{Comte-Bellot}},
  \bibinfo{journal}{Annual review of fluid mechanics}
  \textbf{\bibinfo{volume}{8}}, \bibinfo{pages}{209} (\bibinfo{year}{1976}).

\bibitem[{\citenamefont{Adrian and Goldstein}(1971)}]{LDVReview}
\bibinfo{author}{\bibfnamefont{R.~J.} \bibnamefont{Adrian}} \bibnamefont{and}
  \bibinfo{author}{\bibfnamefont{R.~J.} \bibnamefont{Goldstein}},
  \bibinfo{journal}{Journal of Physics E: Scientific Instruments}
  \textbf{\bibinfo{volume}{4}}, \bibinfo{pages}{505} (\bibinfo{year}{1971}).

\bibitem[{\citenamefont{Wills}(1962)}]{HotWireNearWall1}
\bibinfo{author}{\bibfnamefont{J.}~\bibnamefont{Wills}},
  \bibinfo{journal}{Journal of Fluid Mechanics} \textbf{\bibinfo{volume}{12}},
  \bibinfo{pages}{388} (\bibinfo{year}{1962}).

\bibitem[{\citenamefont{Bhatia et~al.}(1982)\citenamefont{Bhatia, Durst, and
  Jovanovic}}]{HotWireNearWall2}
\bibinfo{author}{\bibfnamefont{J.~C.} \bibnamefont{Bhatia}},
  \bibinfo{author}{\bibfnamefont{F.}~\bibnamefont{Durst}}, \bibnamefont{and}
  \bibinfo{author}{\bibfnamefont{J.}~\bibnamefont{Jovanovic}},
  \bibinfo{journal}{Journal of Fluid Mechanics} \textbf{\bibinfo{volume}{122}},
  \bibinfo{pages}{411} (\bibinfo{year}{1982}).

\bibitem[{\citenamefont{Stefanus et~al.}(2011)\citenamefont{Stefanus, Steers,
  and Goldburg}}]{dirturbsh}
\bibinfo{author}{\bibfnamefont{S.}~\bibnamefont{Stefanus}},
  \bibinfo{author}{\bibfnamefont{S.}~\bibnamefont{Steers}}, \bibnamefont{and}
  \bibinfo{author}{\bibfnamefont{W.}~\bibnamefont{Goldburg}},
  \bibinfo{journal}{Physica D: Nonlinear Phenomena}
  \textbf{\bibinfo{volume}{240}}, \bibinfo{pages}{1873} (\bibinfo{year}{2011}).

\bibitem[{\citenamefont{Chowdhury et~al.}(1984)\citenamefont{Chowdhury,
  Sorensen, Taylor, Merklin, and Lester}}]{brownian}
\bibinfo{author}{\bibfnamefont{D.~P.} \bibnamefont{Chowdhury}},
  \bibinfo{author}{\bibfnamefont{C.~M.} \bibnamefont{Sorensen}},
  \bibinfo{author}{\bibfnamefont{T.}~\bibnamefont{Taylor}},
  \bibinfo{author}{\bibfnamefont{J.}~\bibnamefont{Merklin}}, \bibnamefont{and}
  \bibinfo{author}{\bibfnamefont{T.}~\bibnamefont{Lester}},
  \bibinfo{journal}{Applied optics} \textbf{\bibinfo{volume}{23}},
  \bibinfo{pages}{4149} (\bibinfo{year}{1984}).

\bibitem[{\citenamefont{Fuller et~al.}(1980)\citenamefont{Fuller, Rallison,
  Schmidt, and Leal}}]{FourBall}
\bibinfo{author}{\bibfnamefont{G.}~\bibnamefont{Fuller}},
  \bibinfo{author}{\bibfnamefont{J.}~\bibnamefont{Rallison}},
  \bibinfo{author}{\bibfnamefont{R.}~\bibnamefont{Schmidt}}, \bibnamefont{and}
  \bibinfo{author}{\bibfnamefont{L.}~\bibnamefont{Leal}},
  \bibinfo{journal}{Journal of Fluid Mechanics} \textbf{\bibinfo{volume}{100}},
  \bibinfo{pages}{555} (\bibinfo{year}{1980}).

\bibitem[{\citenamefont{Chu}(1974)}]{ChuLaser}
\bibinfo{author}{\bibfnamefont{B.}~\bibnamefont{Chu}},
  \emph{\bibinfo{title}{Laser light scattering}}
  (\bibinfo{publisher}{Elsevier}, \bibinfo{year}{1974}).

\bibitem[{\citenamefont{Goldburg}(1999)}]{Dynamic}
\bibinfo{author}{\bibfnamefont{W.}~\bibnamefont{Goldburg}},
  \bibinfo{journal}{American Journal of Physics} \textbf{\bibinfo{volume}{67}},
  \bibinfo{pages}{1152} (\bibinfo{year}{1999}).

\bibitem[{\citenamefont{Kataoka et~al.}(1997)\citenamefont{Kataoka, Shibayama,
  Ohuchi, and Yokokawa}}]{Coupling}
\bibinfo{author}{\bibfnamefont{K.}~\bibnamefont{Kataoka}},
  \bibinfo{author}{\bibfnamefont{Y.}~\bibnamefont{Shibayama}},
  \bibinfo{author}{\bibfnamefont{M.}~\bibnamefont{Ohuchi}}, \bibnamefont{and}
  \bibinfo{author}{\bibfnamefont{S.}~\bibnamefont{Yokokawa}},
  \bibinfo{journal}{Applied optics} \textbf{\bibinfo{volume}{36}},
  \bibinfo{pages}{6294} (\bibinfo{year}{1997}).

\bibitem[{\citenamefont{Dufty}(1984)}]{DiffusionShear}
\bibinfo{author}{\bibfnamefont{J.~W.} \bibnamefont{Dufty}},
  \bibinfo{journal}{Physical Review A} \textbf{\bibinfo{volume}{30}},
  \bibinfo{pages}{1465} (\bibinfo{year}{1984}).

\bibitem[{\citenamefont{Tong et~al.}(1988)\citenamefont{Tong, Goldburg, Chan,
  and Sirivat}}]{pairwiseonly}
\bibinfo{author}{\bibfnamefont{P.}~\bibnamefont{Tong}},
  \bibinfo{author}{\bibfnamefont{W.}~\bibnamefont{Goldburg}},
  \bibinfo{author}{\bibfnamefont{C.}~\bibnamefont{Chan}}, \bibnamefont{and}
  \bibinfo{author}{\bibfnamefont{A.}~\bibnamefont{Sirivat}},
  \bibinfo{journal}{Physical Review A} \textbf{\bibinfo{volume}{37}},
  \bibinfo{pages}{2125} (\bibinfo{year}{1988}).

\bibitem[{\citenamefont{Einstein}(1905)}]{einsteinBrownian}
\bibinfo{author}{\bibfnamefont{A.}~\bibnamefont{Einstein}},
  \bibinfo{journal}{Annalen der Physik} \textbf{\bibinfo{volume}{17}},
  \bibinfo{pages}{16} (\bibinfo{year}{1905}).

\bibitem[{\citenamefont{Macosko and Larson}(1994)}]{rheology}
\bibinfo{author}{\bibfnamefont{C.~W.} \bibnamefont{Macosko}} \bibnamefont{and}
  \bibinfo{author}{\bibfnamefont{R.~G.} \bibnamefont{Larson}},
  \emph{\bibinfo{title}{Rheology: principles, measurements, and applications}}
  (\bibinfo{publisher}{VCH New York}, \bibinfo{year}{1994}).

\bibitem[{\citenamefont{Snyder and Lumley}(1971)}]{snyder1971}
\bibinfo{author}{\bibfnamefont{W.~H.} \bibnamefont{Snyder}} \bibnamefont{and}
  \bibinfo{author}{\bibfnamefont{J.}~\bibnamefont{Lumley}},
  \bibinfo{journal}{Journal of Fluid Mechanics} \textbf{\bibinfo{volume}{48}},
  \bibinfo{pages}{41} (\bibinfo{year}{1971}).

\bibitem[{\citenamefont{Sdougos et~al.}(1984)\citenamefont{Sdougos, Bussolari,
  and Dewey}}]{C&Psecondary}
\bibinfo{author}{\bibfnamefont{H.}~\bibnamefont{Sdougos}},
  \bibinfo{author}{\bibfnamefont{S.}~\bibnamefont{Bussolari}},
  \bibnamefont{and} \bibinfo{author}{\bibfnamefont{C.}~\bibnamefont{Dewey}},
  \bibinfo{journal}{Journal of Fluid Mechanics} \textbf{\bibinfo{volume}{138}},
  \bibinfo{pages}{379} (\bibinfo{year}{1984}).

\end{thebibliography}
\end{document}